\documentclass{Pos}

\newcommand{\be}{\begin{equation}}
\newcommand{\ee}{\end{equation}}
\newcommand{\bea}{\begin{eqnarray}}
\newcommand{\eea}{\end{eqnarray}}

\title{Confinement and chiral symmetry breaking, one or two 
critical points?}

\ShortTitle{QCD critical points}

\author{\speaker{Pedro Bicudo}\thanks{I thank the organizers for maintaining this beautiful meeting in the spirit of Ileana Iori.}\\
        CFTP, Dep. F\'{\i}sica, Instituto Superior T\'ecnico,
Av. Rovisco Pais, 1049-001 Lisboa, Portugal\\
        E-mail: \email{bicudo@ist.utl.pt}}

\abstract{We study the QCD phase diagram, in particular we study the critical points of the
two main QCD phase transitions, confinement and chiral symmetry breaking.
Confinement drives chiral symmetry breaking, and, due to the finite quark mass,
at small density both transitions are a crossover, while they are a first or
second order phase transition in large density. We study the QCD phase diagram
with a quark potential model including both confinement and chiral symmetry. This
formalism, in the Coulomb gauge hamiltonian formalism of QCD, is presently the
only one able to microscopically include both a quark-antiquark confining potential and
a vacuum condensate of quark-antiquark pairs. 
Our order parameters are the Polyakov loop and the quark mass gap. 
The confining potential is extracted from the Lattice
QCD data of the Bielefeld group. We scan the QCD phase diagram for different
quark masses, in order to address how the quark masses affect the critical point
location in the phase diagram.}

\FullConference{XLVIII International Winter Meeting on Nuclear Physics, BORMIO2010\\
		January 25-29, 2010\\
		Bormio, Italy}

\begin{document}

\section{Motivation}

Our main motivation is to contribute to understand the QCD phase
diagram, for finite $T$ and $\mu$, to be studied at LHC, RHIC and FAIR.

Using modern quark models for light quarks we now
study chiral symmetry breaking, i. e. quark mass generation, at finite T.
In the last Bormio Meeting we used the bottomonium and charmonium as good
prototypes to study finite T quark-antiquark potentials. In that case it was sufficient
to solve the Schrödinger equation with static lattice QCD potentials, since
$
m_b, m_c >> \Lambda_{QCD} 
$
and
$
m_b, m_c  >> Tc
$
allowed us to neglect in the quark sector, temperature effects, spontaneous
chiral symmetry breaking, relativistic effects and coupled channels. 
However,
to address light quarks at finite temperature $T$ it is also necessary to include
temperature, not only in the confining quark-antiquark interaction, but also
in the quark mass generation and in the quark propagator. 

Thus the present work, not only addresses the QCD phase diagram,
but it also constitutes the first step to allow us in the future to,
\\ - compute the spectrum of any hadron at finite T ,
\\ - compute the interaction of any hadron-hadron system at finite T .

\begin{figure}[t]
\hspace{-2.5cm}
\center{
\includegraphics[width=0.6\columnwidth]{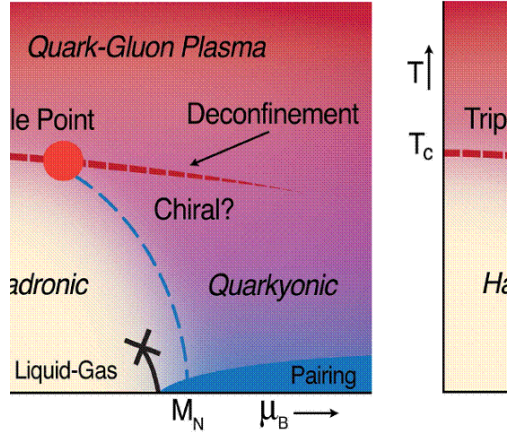}
 }
\hspace{-1.5cm} 
\caption{A sketch of the QCD Phase Diagram, according to
the collaboration CBM at FAIR
\cite{CBM}.  It is usually assumed that the critical point for deconfinement
coincides with the critical point for chiral symmetry restoration.}
\label{CBM}
\end{figure}

Here we address the finite temperature string tension, the quark mass gap
for a finite current quark mass and temperature, and the deconfinement and
chiral restoration crossovers. We conclude on the separation of the critical 
point for for chiral symmetry restoration from the critical point for
deconfinement.

\section{Fits for the finite T string tension from the Lattice QCD energy $F_1$}

At vanishing temperature $T=0$, the confinement, modelled by a string, is dominant at moderate distances,
\be
V(r) \simeq  {\pi \over 12 r} + V_0 + \sigma  r \ .
\ee 
At short distances we have the Luscher or Nambu-Gotto Coulomb due to the
string vibration + the OGE coulomb, however the Coulomb is not important for
chiral symmetry breaking. At finite temperature the string tension $\sigma(T)$
should also dominate chiral symmetry breaking, and thus one of our crucial steps 
here is the fit of the string tension $\sigma(T)$
from the Lattice QCD data of the Bielefeld Lattice QCD group,
\cite{Doring:2007uh,Hubner:2007qh,Kaczmarek:2005ui,Kaczmarek:2005gi,Kaczmarek:2005zp}.

\begin{figure}[t!]
\hspace{0cm}
\center{
\includegraphics[width=0.45\columnwidth]{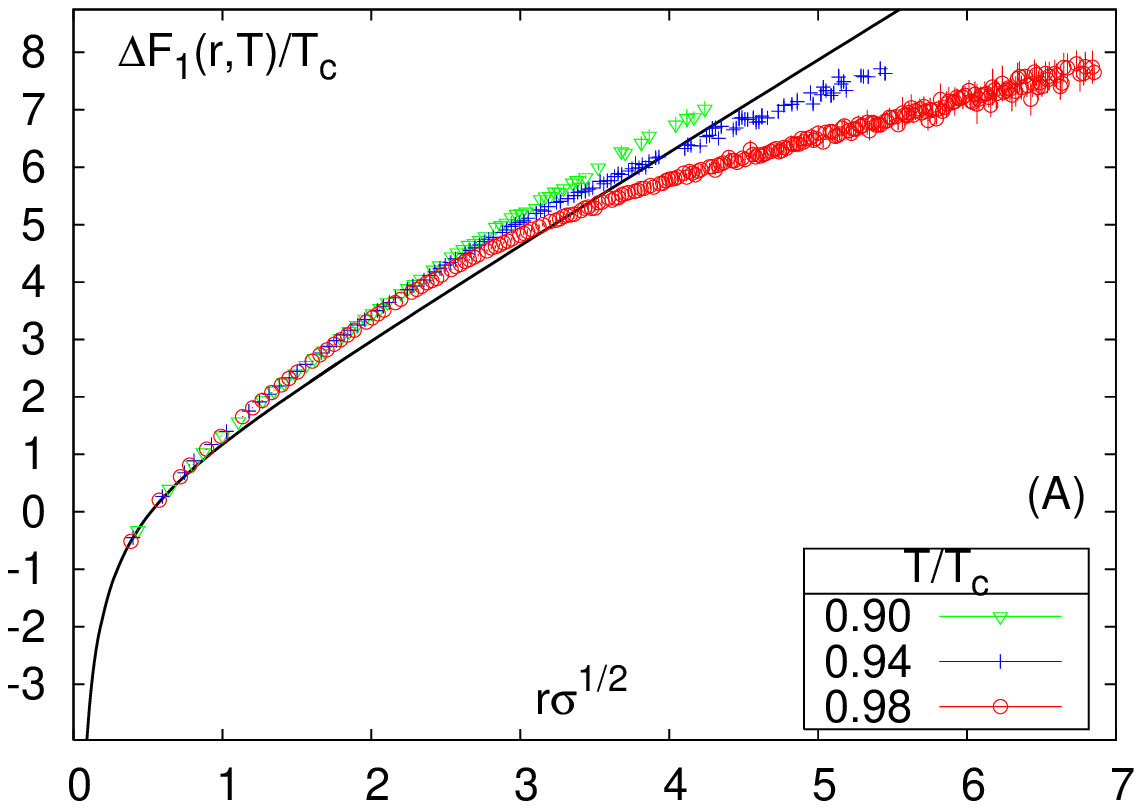}
\hspace{1cm}
\includegraphics[width=0.35\columnwidth]{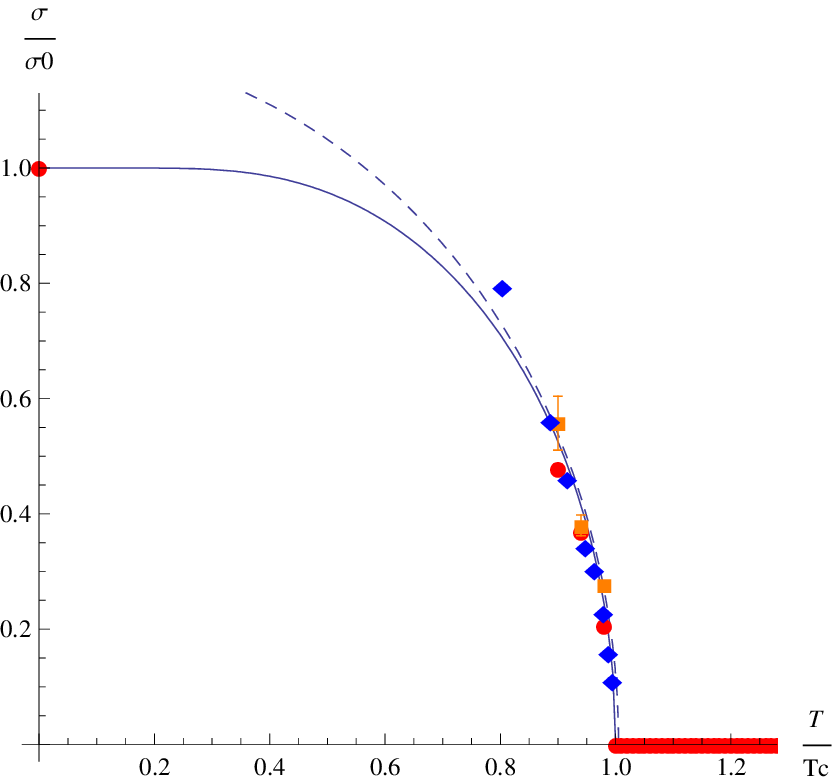}
}
\caption{
Left: The Bielefeld fee $F_1$ energy at $T<T_c$.
Right: comparing the magnetization critical curve with the string tension 
$\sigma / \sigma_0$, fitted from the long distance part of $F_1$, 
they are quite close.
}
\label{magnetiz}
\end{figure}

The Polyakov loop is a gluonic path, closed in the imaginary time $t_4$ (proportional
to the inverse temperature $T^{-1}$) direction in QCD discretized 
in a periodic boundary euclidian Lattice. 
It measures the free energy $F$ of one or more static quarks,
\be
P(0) = N e^{ - F_q /T } \ , \ \ P^a(0)\bar P^{\bar a}(r) = N e^{ - F_{ q \bar q}(r)  /T }  \ .
\ee
If we consider a single solitary quark in the universe, in the confining
phase, his string will travel as far as needed to connect the quark to an antiquark,
resulting in an infinite energy F. Thus the 1 quark Polyakov loop $P$ is a
frequently used order parameter for deconfinement.
With the string tension $\sigma(T)$ 
extracted from the $q \bar q$ pair of Polyakov loops
we can also estimate the 1 quark Polyakov loop $P(0)$.
At finite $T$, we use as thermodynamic
potentials the free energy $F_1$ and the
internal energy $U_1$, computed in Lattice
QCD with the Polyakov loop
\cite{Doring:2007uh,Hubner:2007qh,Kaczmarek:2005ui,Kaczmarek:2005gi,Kaczmarek:2005zp}.
They are related to the static potential
$V(r)   = - f d r$
with
$F1 (r)= - f d r - S d T$
adequate for isothermic transformations.
In Fig. \ref{magnetiz} we extract the string tensions $\sigma(T)$
from the free energy $F_1(T)$ 
computed by the Bielefeld group, and we also include string tensions
previously computed by the Bielefeld group 
\cite{Kaczmarek:1999mm}.

We also find an ansatz for the string tension curve, among
the order parameter curves of other physical systems related to 
confinement, i. e. in ferromagnetic materials, in the Ising
model, in superconductors either in the BCS model or in the
Ginzburg-Landau model, or in string models, to suggest 
ansatze for the string tension curve. We find that the order parameter
curve that best fits our string tension curve is
the spontaneous magnetization of a ferromagnet 
\cite{FeynmanLS},
solution of the
algebraic equation,
\be
{M \over M_{sat}} = \tanh \left(  { T_c \over T}  {M \over M_{sat}}  \right) \ .
\label{eqformagnetization}
\ee
In Fig. \ref{magnetiz} we show the solution of Eq. \ref{eqformagnetization}
obtained with the fixed point expansion, and compare it with the
string tensions computed from lattice QCD data.

\section{The mass gap equation with finite $T$ and
finite current quark mass $m_0$.}

\begin{figure}[t!]
\includegraphics[width=0.45\columnwidth]{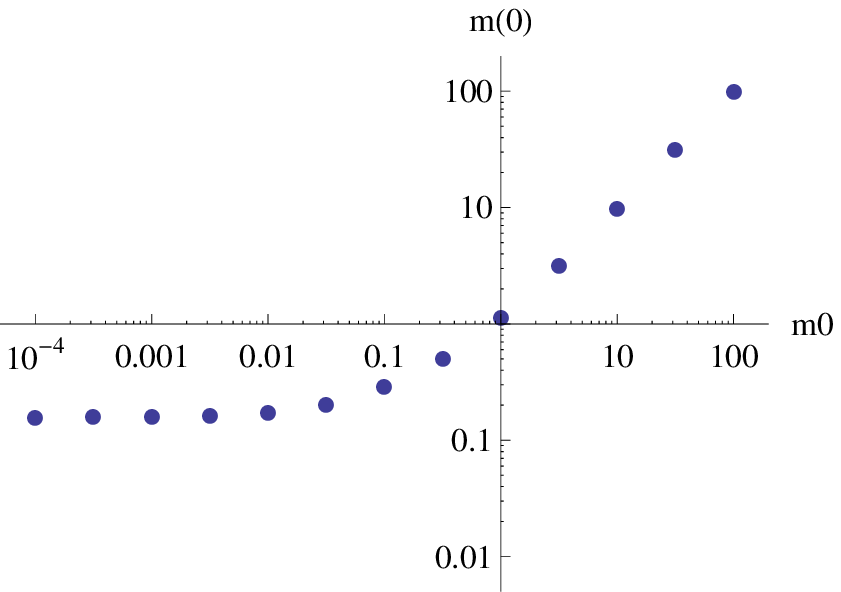}
\includegraphics[width=0.5\columnwidth]{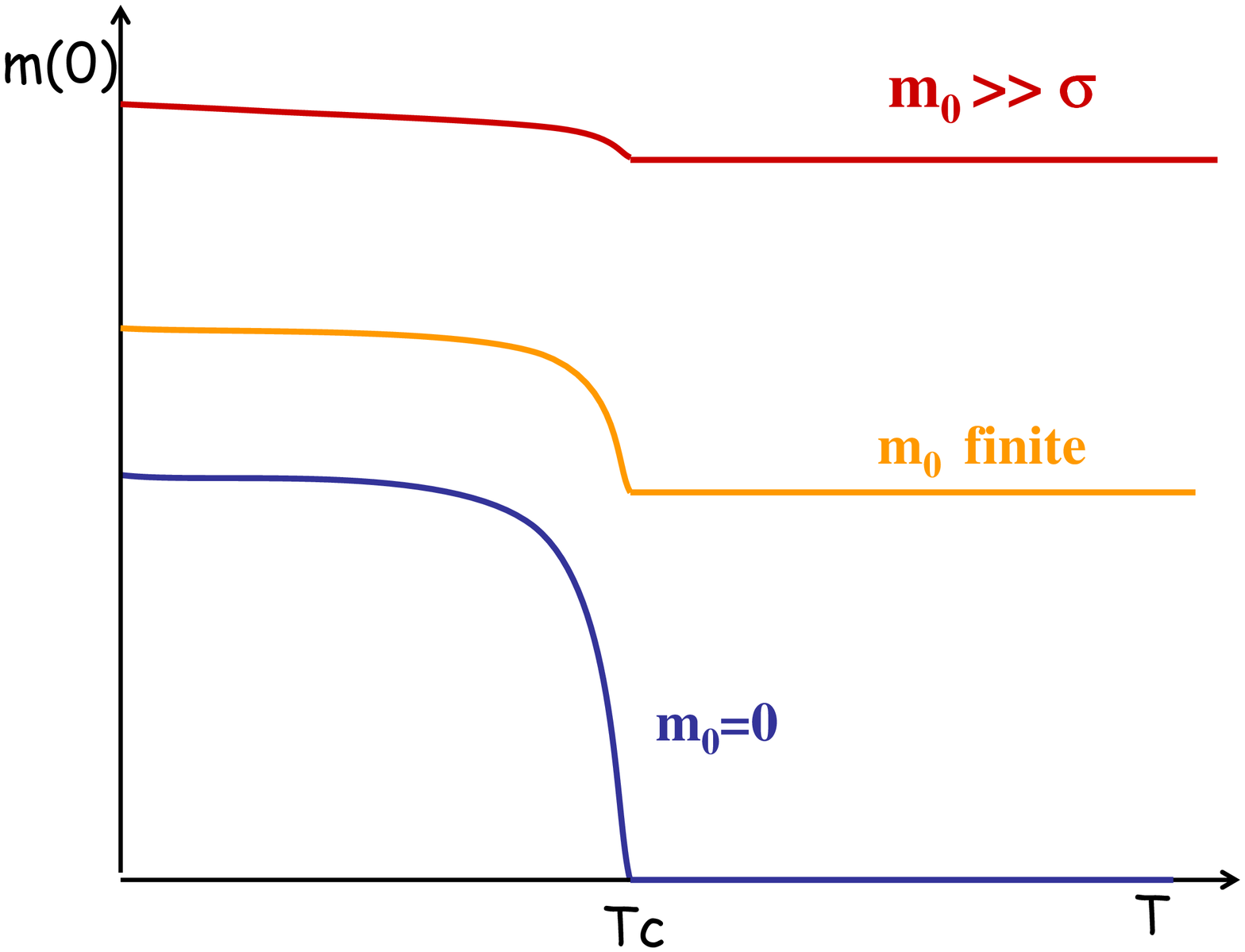}
\caption{
Left: the mass gap $m(0)$ solution of 
as a function of the quark current mass $m_0$, in units of $\sigma=1$. 
Right: sketch of the effect of $m_0$ on the crossover versus phase transition of
choral restoration at finite $T$.
}
\label{massgap}
\end{figure}

Now, the critical point occurs when the phase transition changes to a crossover,
and the crossover in QCD is produced by the finite current quark mass m0,
since it affects the order parameters $P$ or $\sigma$, and  the mass gap $m(0)$
 or the quark condensate $\langle \bar q q \rangle$.
The mass gap equation at the ladder/rainbow truncation of Coulomb
Gauge QCD in equal time reads,
\bea
m(p) &=& m_0+ { \sigma \over p^3}
\int_0^\infty {d k \over 2 \pi}  {1 \over \sqrt{k^2 + m(k)^2}}  \biggl\{ \left[
{ p k \over (p-k)^2 + \mu^2} 
- { p k \over (p+k)^2 + \mu^2} \right]
 m (k) p  
\nonumber \\
&&  - \left[
{ p k \over (p-k)^2 + \mu^2} 
+ { p k \over (p+k)^2 + \mu^2} + {1 \over 2} \log  {(p-k)^2 + \mu^2 \over (p+k)^2 + \mu^2}\,
 \right]
m(p) k  \biggr\}  \ .
\label{fixedpointeq}
\eea
The mass gap equation (\ref{fixedpointeq})
for the running mass
$m(p)$ is a non-linear
integral equation with
a nasty cancellation
of Infrared divergences
\cite{Adler:1984ri, Bicudo:2003cy,
LlanesEstrada:1999uh}.
We devise a new method
with a rational ansatz,
and with relaxation,
to get a maximum
precision in the IR
where the equation is
nearly almost unstable.
The solution $m(p)$  is shown in Fig. \ref{massgap} for a vanishing momentum
$p=0$.

At finite $T$, one only has to change the string tension to the finite T
string tension $\sigma(T)$
\cite{Bicudo:2010hg}, 
and also to replace an integral in $\omega$ 
by a sum in Matsubara Frequencies. Both are equivalent to a reduction in the
string tension, $\sigma \to \sigma^*$ and thus all we have to do is to solve the mass gap
equation in units of $\sigma^*$ .
The results are depicted in Fig. \ref{massgap}.
Thus at vanishing $m_0$ we have a chiral symmetry phase transition,
and at finite $m_0$ we have a crossover,
that gets weaker and weaker when $m_0$ increases. This is also sketched
in Fig. \ref{massgap}.

\section{Chiral symmetry and confinement
crossovers with a finite current quark mass}

\begin{figure}[t!]
\hspace{0cm}
\includegraphics[width=0.5\columnwidth]{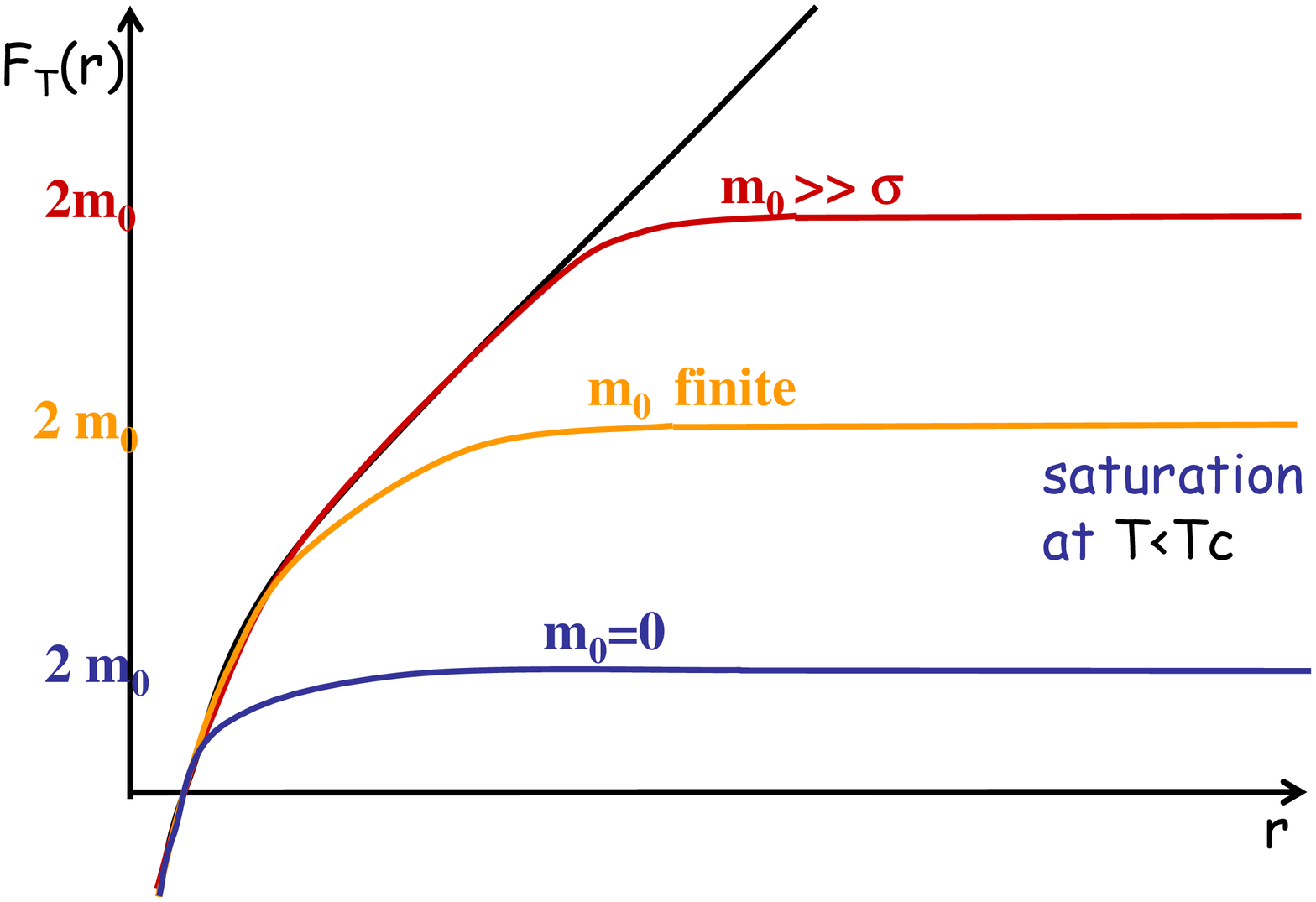}
\includegraphics[width=0.55\columnwidth]{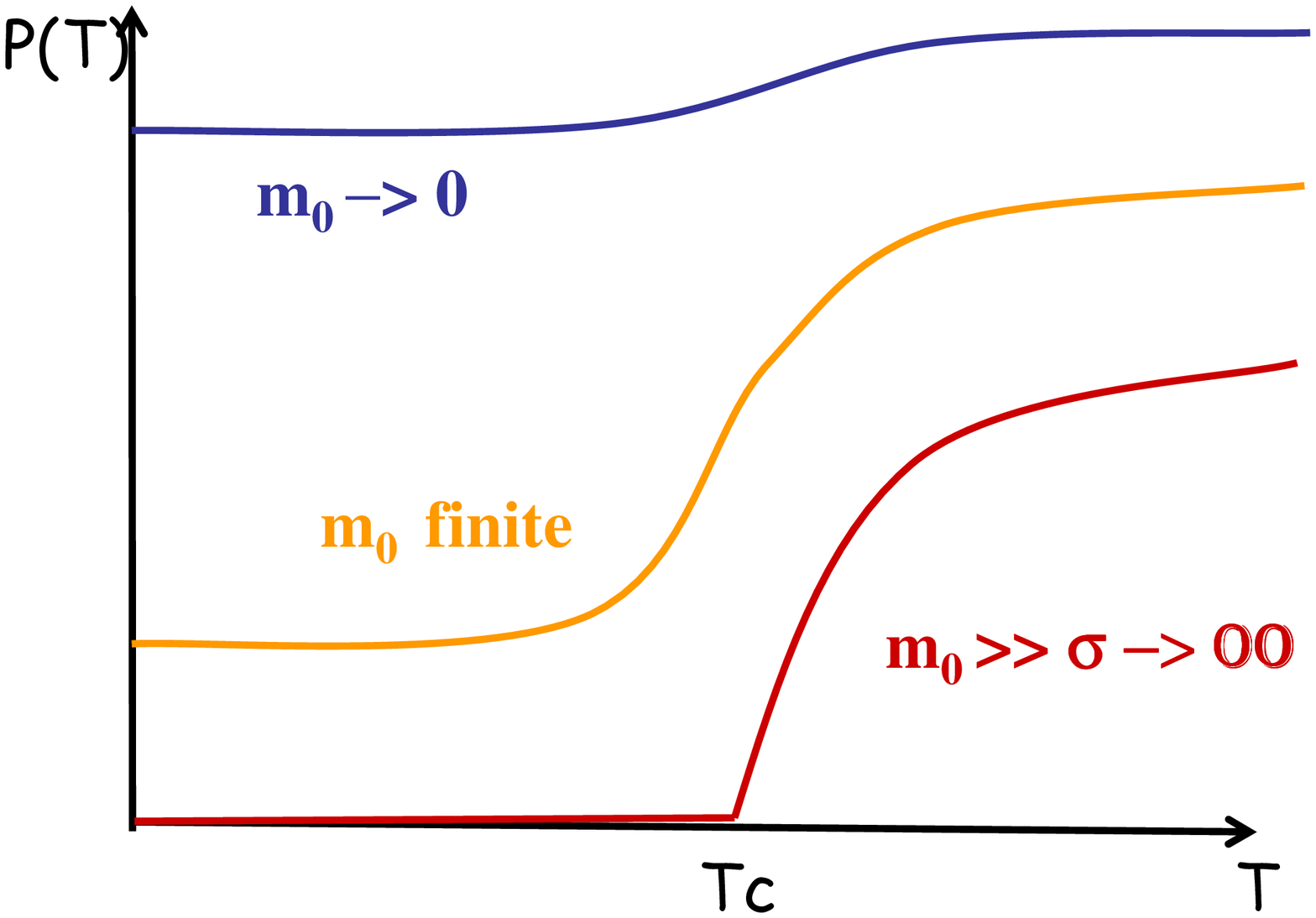}
\vspace{-1cm}
\caption{Sketches of the saturation of confinement (left), and of the corresponding 
crossover in the order parameter $P$ polyakov loop (righ). 
}
\label{sturationconfinement}
\end{figure}

In what concerns confinement, the linear confining quark-antiquark potential
saturates when the string breaks at the threshold for the creation of 
a quark-antiquark pair.
Thus the free energy $F(0)$ of a single static quark is not infinite, 
but is the energy of the string saturation, 
of the order of the mass of a meson i. e. of $ 2 m_0$.
For the Polyakov loop we get,
\be
P(0) \simeq N e^{ - 2 m_0 / T} \ .
\ee
Thus at infinite $m_0$ we have a confining phase transition,
while at finite $m_0$ we have a crossover,
that gets weaker and weaker when $m_0$ decreases.
This is sketched in Fig. \ref{sturationconfinement}.

Since the finite current quark mass affects in opposite ways the crossover
for confinement and the one for chiral symmetry, we conjecture that at finite $T$ and $\mu$
there are not only one but two critical points (a point where a crossover
separates from a phase transition). 
Since for the light $u$ and $d$ quarks 
the current mass $m_0$ is small, we expect the crossover for chiral symmetry restoration 
critical to
be closer to the $\mu=0$ vertical axis, and the crossover for deconfinement
to go deeper into the finite $\mu$ region of the critical curve in the QCD
phase diagram depicted in Fig. \ref{CBM}.


\begin{thebibliography}{99}
\bibitem{CBM} 
CBM Progress Report, publicly available at
http://www.gsi.de/fair/experiments/CBM, (2009);
%
  J.~M.~Heuser  [CBM Collaboration],
  Nucl.\ Phys.\  A {\bf 830}, 563C (2009)
  [arXiv:0907.2136 [nucl-ex]].
  
  
  
\bibitem{Doring:2007uh}
  M.~Doring, K.~Hubner, O.~Kaczmarek and F.~Karsch,
  Phys.\ Rev.\  D {\bf 75}, 054504 (2007)
  [arXiv:hep-lat/0702009].

\bibitem{Hubner:2007qh}
  K.~Hubner, F.~Karsch, O.~Kaczmarek and O.~Vogt,
  arXiv:0710.5147 [hep-lat].
  
\bibitem{Kaczmarek:2005ui}
  O.~Kaczmarek and F.~Zantow,
  Phys.\ Rev.\  D {\bf 71}, 114510 (2005)
  [arXiv:hep-lat/0503017].
  
\bibitem{Kaczmarek:2005gi}
  O.~Kaczmarek and F.~Zantow,
  arXiv:hep-lat/0506019.

\bibitem{Kaczmarek:2005zp}
  O.~Kaczmarek and F.~Zantow,
  PoS {\bf LAT2005}, 192 (2006)
  [arXiv:hep-lat/0510094].
  

  
\bibitem{Kaczmarek:1999mm}
  O.~Kaczmarek, F.~Karsch, E.~Laermann and M.~Lutgemeier,
  Phys.\ Rev.\  D {\bf 62}, 034021 (2000)
  [arXiv:hep-lat/9908010].
  
\bibitem{FeynmanLS}
  R.~Feynamn, R.~Leighton, M.~Sands,
  "The Feynman Lectures on Physics", Vol II, chap. 36 "Ferromagnetism",
 published by Addison Wesley Publishing Company, Reading, Massachussets, ISBN 0-201-02117-x (1964).
 
\bibitem{Adler:1984ri}
  S.~L.~Adler and A.~C.~Davis,
  Nucl.\ Phys.\  B {\bf 244}, 469 (1984).
  
\bibitem{Bicudo:2003cy}
  P.~J.~A.~Bicudo and A.~V.~Nefediev,
  Phys.\ Rev.\  D {\bf 68}, 065021 (2003)
  [arXiv:hep-ph/0307302].
  
\bibitem{LlanesEstrada:1999uh}
  F.~J.~Llanes-Estrada and S.~R.~Cotanch,
  Phys.\ Rev.\ Lett.\  {\bf 84}, 1102 (2000)
  [arXiv:hep-ph/9906359].

\bibitem{Bicudo:2010hg}
  P.~Bicudo,
  arXiv:1003.0936 [hep-lat].

\end{thebibliography}
\end{document}